\documentclass[fleqn,usenatbib]{mnras}

\usepackage{newtxtext,newtxmath}
\usepackage{arydshln}

\usepackage[T1]{fontenc}
\DeclareMathOperator*{\dprime}{\prime \prime}
\newcommand{\xmm}{{\it XMM-Newton }}
\newcommand{\nustar}{\textit{NuSTAR }}
\newcommand{\code}[1]{{\texttt{#1}}}
\DeclareRobustCommand{\VAN}[3]{#2}
\let\VANthebibliography\thebibliography
\def\thebibliography{\DeclareRobustCommand{\VAN}[3]{##3}\VANthebibliography}

\usepackage{graphicx}	
\usepackage{amsmath}	

\title[SXP~15.3 \& SXP~305]{Disentangling the neighbouring pulsars SXP~15.3 and SXP~305}

\author[Monageng et al.]
  {I. M. Monageng$^{1,3}$\thanks{E-mail: itu@saao.ac.za}, 
  M. J. Coe$^{2}$,
  L. J. Townsend$^{1,5}$,
    S. G. T. Laycock$^{6}$,
    J.~A. Kennea$^{7}$,
  \newauthor
   A. Roy$^{6}$, A. Udalski$^{4}$, S. Bhattacharya$^{6}$, D. M. Christodoulou$^{6}$, D. A. H. Buckley$^{1,3,9}$
  \newauthor
     and P. A. Evans$^{8}$
\\
$^1$South African Astronomical Observatory, P.O Box 9, Observatory, 7935, Cape Town, South Africa\\
$^2$Physics \& Astronomy, University of Southampton, SO17 1BJ, UK\\
$^3$Department of Astronomy, University of Cape Town, Private Bag X3, Rondebosch 7701, South Africa\\
$^4$Astronomical Observatory, University
of Warsaw, Al. Ujazdowskie 4, 00-478 Warszawa, Poland\\
$^5$Southern African Large Telescope, P.O. Box 9, Observatory, 7935, Cape Town, South Africa\\
$^{6}$ Lowell Center for Space Science and Technology, University of Massachusetts Lowell, MA, 01854, USA \\
$^{7}$ Department of Astronomy and Astrophysics, The Pennsylvania State University, 525 Davey Lab, University Park, PA 16802, USA \\
$^{8}$ University of Leicester, X-ray and Observational Astronomy Research Group, Leicester Institute for Space and Earth Observation \\
$^{9}$ Department of Physics, University of the Free State, PO Box 339, Bloemfontein, 9300, South Africa\\
}

\date{Accepted XXX. Received YYY; in original form ZZZ}

\pubyear{2015}

\begin{document}
\label{firstpage}
\pagerange{\pageref{firstpage}--\pageref{lastpage}}
\maketitle

\begin{abstract}
SXP~15.3 and SXP~305 are two Be X-ray binaries in the Small Magellanic Cloud that are spatially separated by $\sim$7 arcsec. The small separation between these sources has, in the past, resulted in confusion about the origin of the emission from the combined region. We present long-term optical and X-ray monitoring results of both sources, where we study the historic and recent behaviour. In particular, from data collected as part of the S-CUBED project we see repeating X-ray outbursts from the combined region of the two sources in the recent lightcurve from the Neil Gehrels \textit{Swift} Observatory, and we investigate the origin of this emission. Using the H$\alpha$ emission line from the Southern African Large Telescope (SALT) and photometric flux from the Optical Gravitational Lensing Experiment (OGLE) to study the changes in the size and structure of the Be disc, we demonstrate that the X-ray emission likely originates from SXP~15.3. Timing analysis reveals unusual behaviour, where the optical outburst profile shows modulation at approximately twice the frequency of the X-ray outbursts. We consider either of these periodicities being the true orbital period in SXP~15.3 and propose models based on the geometric orientations of the Be disc and neutron star to explain the physical origin of the outbursts.

\end{abstract}




\section{Introduction}

\begin{figure}
\resizebox{\hsize}{!}
            {\includegraphics[angle=0,width=12cm]{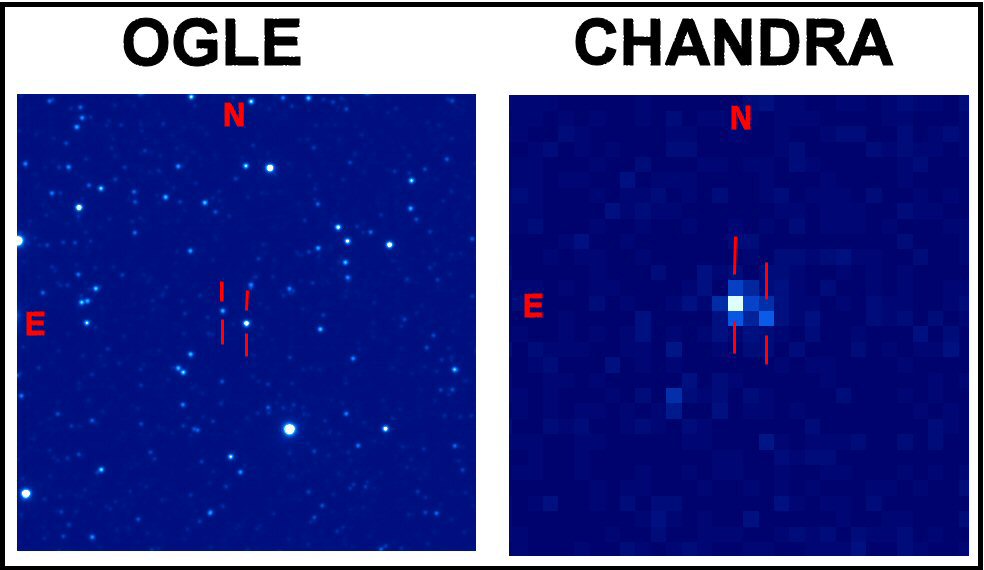}}
\caption{Optical (left) and X-ray (right) images of the 2 x 2 arcmin field containing SXP~15.3 (south-west) and SXP~305 (north-east). The optical image is an OGLE IV I-band image taken on 2010 September 14. The X-ray image is from Chandra ACIS-I from 2006 November 22. The two sources are separated by 6.87 arcsec.}
\label{fig:double}
\end{figure}

Be X-ray binaries (BeXBs) make up the largest fraction of the high mass X-ray X-ray binary systems (HMXBs), with 49\% of the total population composed of them \citep{2013ApJ...764..185C}. They primarily comprise a neutron star (NS) in an eccentric orbit around a massive B-spectral type star with a circumstellar disc. On rare occasions the compact object orbiting the Be star is a black hole  \citep{2014Natur.505..378C} or white dwarf \citep{2020MNRAS.tmpL.106C}. The interaction between the NS and Be star can result in enhanced X-ray emission due to the accretion of matter from the circumstellar disc by the NS. This transient X-ray behaviour is classically defined by two types of outbursting events, Type I and Type II. Type I outbursts have luminosities on the order of $10^{37}$~erg. s$^{-1}$ or less, and are typically associated with periastron passage \citep{1986ApJ...308..669S}. Type II outbursts are typically an order of magnitude brighter than Type I outbursts and are not associated with a specific orbital phase. The origin of Type II outbursts have proved elusive, with various explanations proposed alluding to complex circumstellar disc variability and/or orbital geometry (e.g. \citealt{2011PASJ...63L..25M, 2014ApJ...790L..34M, 2017MNRAS.464..572M}). Optical and infrared observations are used to trace the circumstellar disc variability, primarily through the changes in the H$\alpha$ emission line. \\
The Small Magellanic Cloud (SMC) hosts a large number of HMXBs (121; \citealt{2016A&A...586A..81H}), which is higher than those in the Milky Way (65; \citealt{2011MNRAS.413.1600R}) despite being $\sim$50 times less massive. This large abundance of HMXBs in the SMC is likely due to a recent occurrence of star formation as a result of an increase in the tidal force from its interaction with the Large Magellanic Cloud \citep{1996MNRAS.278..191G, 2011MNRAS.413.2015D}. \\
In this manuscript we present analysis performed on two sources located in the SMC, SXP~15.3 and SXP~305, which are separated spatially by just $\sim$7 arcsec - see Fig ~\ref{fig:double}. 

The close proximity of the two objects has resulted in confusion about their optical counterparts, as they both fall within the X-ray error circles from most missions \citep{2000A&AS..142...41H, 2001A&A...374.1009C, 2011MNRAS.412..391S}. \cite{2001A&A...374.1009C} presented the most accurate X-ray positions using \textit{ROSAT HRI} data and confirmed the association of SXP~15.3 with the B-spectral type star listed as [MA93]552 in the optical catalogues of \cite{1993A&AS..102..451M}. These authors use medium resolution spectroscopy to classify the optical counterpart as OIIIe or B0Ve. The optical counterpart of SXP~305 was noted to show no emission lines and to be a later B-spectral type than [MA93]552. The 15.3~s pulse period of SXP~15.3 was derived by \cite{1999IAUC.7081....4L} through the analysis of \textit{ROSAT} and \textit{BATSE} data. The detection of the 305~s pulse period of SXP~305 has for a long time proved elusive, having only been detected recently using \textit{NuSTAR} observations \citep{2019ApJ...884....2L}. \\
In this paper we investigate the origin of recent outbursting behaviour coming from the combined field of both sources as seen from \textit{Swift XRT} observations. We use optical photometry and spectroscopy to study the long-term variability of the Be discs from both objects, where we see them exhibiting contrasting behaviour.

\section{Observations}
\subsection{OGLE}
Long-term $I-$band photometry monitoring with a typical cadence of 1-3 days is performed with the Optical Gravitational Lensing Experiment (OGLE) project \citep{1997AcA....47..319U,2015AcA....65....1U}. The optical counterpart of SXP~15.3 has been observed for $\sim$23 years in the $I-$band, where it was identified as SMC-SC6-99923 in OGLE II, SMC100.1 43671 in OGLE III and SMC720.26 47 in OGLE IV. The optical counterpart of SXP~305 has also been observed in the same period with identifiers SMC-SC6-99991 in OGLE II, SMC100.1 43700 in OGLE III and SMC720.26 531 in OGLE IV. Both objects have also been observed in the $V-$band with a lower cadence. The standard OGLE pipeline \citep{2015AcA....65....1U} was used for data reduction and calibrations.

\subsection{SALT}
We observed the optical counterpart of SXP~15.3 using the Robert Stobie Spectrograph (RSS; \citealt{2003SPIE.4841.1463B,2003SPIE.4841.1634K}) on the Southern African Large Telescope (SALT; \citealt{2006SPIE.6267E..0ZB}) between September 2017 and October 2019. The PG1800 grating was used to cover the wavelength region that includes the H$\alpha$ line ($\sim$6000$-7250$~\AA). A single observation covering the blue wavelength region ($\sim$3860$-$4900\AA) was obtained using the PG2300 grating.

The optical counterpart of SXP~305 was observed twice with the PG1800 grating and once with the PG2300 grating. The close proximity of the two objects allowed for simultaneous observations to be performed on them by arranging the slit angle accordingly.

The primary reductions, which include overscan correlation, bias subtraction, gain and amplifier cross-talk corrections were performed using the SALT pipeline \citep{2012ascl.soft07010C}. We then used \textsc{iraf}\footnote{Image Reduction and Analysis Facility: iraf.noao.edu} for the remaining steps: wavelength calibrations, background subtraction and extraction of the one-dimensional spectra. 

\subsection{Swift}
The S-CUBED survey \citep{kennea2018} is a shallow weekly survey of the optical extent of the SMC by the Swift X-ray Telescope (XRT; \citealt{burrows05}). Individual exposures in the S-CUBED survey are typically 60s long, and occur weekly, although interruptions can occur due to scheduling constraints. An automated pipeline based on the tools of \cite{2009MNRAS.397.1177E} was used to analyse the data. The full details of the data reduction steps are given in \cite{kennea2018}. The sources that are the subject of this paper have been observed since June 2016 with S-CUBED.

\subsection{RXTE}

The X-ray
monitoring of the SMC was carried out for over a decade using the Rossi X-ray Timing Explorer space observatory (RXTE) \citep{jahoda1996J}. The
Proportional Counter Array (PCA) instrument on board
RXTE was observing the SMC 1997 - 2012, full details of
these observations and analysis are found in \cite{laycock2005} and \cite{galache2008}. Positional identification of
SMC HMXBs is impossible with RXTE due to its 2 degree
full
width at zero intensity field-of-view. Identification is performed solely through pulse timing analysis and
identification with related activity in optical light curves. 
RXTE data on SXP 15.3 were initially presented in \cite{2011MNRAS.412..391S}.

\subsection{Chandra}

\textit{Chandra} was launched in 1999 and resides in a highly elliptical 64-hour orbit. It contains an assembly of grazing-incidence mirrors which focusses the photons onto two scientific instruments namely the Advanced CCD Imaging Spectrometer (ACIS) or the HighResolution Camera (HRC). The observations were processed through standard \code{CIAO} data analysis threads. The data was barycenter corrected and \code{fluximage} was run in the broad energy range (0.5-7.0 keV), with creation of exposure maps. The point sources were detected by running \code{wavedetect} with eavelet scales of 2,4,8 \& 16. The source and background light curves were extracted in the broad energy band binned at 3.25 sec, the native resolution for \textit{Chandra}. For extraction the source region was defined as a circle centered on the source coordinates with a 95\% encircled-energy radius at 2.3 keV as determined by the \code{evalpos} tool applied to the PSF map for that observation.  The background region was an annulus with an inner radius of twice the source region radius and an outer radius of five times the source region radius, also centered on the source coordinates.

\subsection{XMM}

The \xmm observatory \citep{Jansen} was launched by the European Space Agency (ESA) in 1999 consisting of three high throughput X-ray telescopes. The three telescopes focusses the photons onto European Photon Imaging Camera (EPIC) and other spectrographic instruments. \xmm looked at the field of view containing the two pulsars four times (ObsIDs: 0110000101,0404680301,0601213001,0601211401). The data is processed through standard tasks from Science Analysis Software (SAS) version 17.0.0. The event files were cleaned for good time intervals using \code{tabgtigen} and barycenter corrected with the \code{barycen} tool. We applied Standard filters to extract MOS (Patterns <= 12) and PN (Patterns <= 4, single and double pixel events) event files. The background subtracted lightcurves were extracted using the command \code{epiclccorr} and binned at 2.7982 s with the source events extracted from a radius of $20^{\dprime}$ and background region from an annulus of inner radius $30^{\dprime}$ and outer radius $60^{\dprime}$  about the detected source coordinate.

\subsection{NuSTAR}

The \textit{Nuclear Spectroscopic Telescope Array}~(\nustar;\citet{harrison2013nuclear}) is the first focusing high-energy telescope in orbit, launched on 2012 June 13. It operates in the 3 - 79 keV energy band, enabling us to overcome the $\sim$ 10keV barrier set by previous X-ray telescopes. \nustar consist of two co-aligned hard X-ray grazing incident telescopes (Wolter I) which focus on two independent solid state focal plane detectors (focal length $\sim$ 10 m). The two focal plane images can be added to gain sensitivity as the optics and detectors are designed as identical as possible. 

All the archival data(listed in Table\ref{tab:obs}) for both FPMA and FPMB have been analyzed using The \nustar Data Analysis Software (NuSTARDAS v1.8.0) and CALDB (v20190513). This module is integrated in the Heasoft (v6.26.1) package distributed by High Energy Archive Science Research Center (HEASARC). Using the \code{nupipeline} command the level 1 event files are converted to level 2 event files. Standard cleaning process has been used. The gti and barycenter corrected high-level products (level3: spectra, lightcurves and skymap images) were created using the \code{nuproducts} command. These products have been created in the broad energy band. Source extraction region of $20^{\dprime}$ was chosen centered at the known SXP 15.3 and SXP 305 coordinates, background extraction was done with a annular region (inner:$30^{\dprime}$, outer: $60^{\dprime}$) centered at the same source coordinate. The net lightcuves, binned at 0.07s were background subtracted using \code{lcmath} for both FPMA and FPMB and they were added to get the total background subtracted light curve using the same command for each source in each observations. 
\section{Results}
\subsection{X-ray variability}
\begin{figure}
    \centering
    \includegraphics[width=9cm]{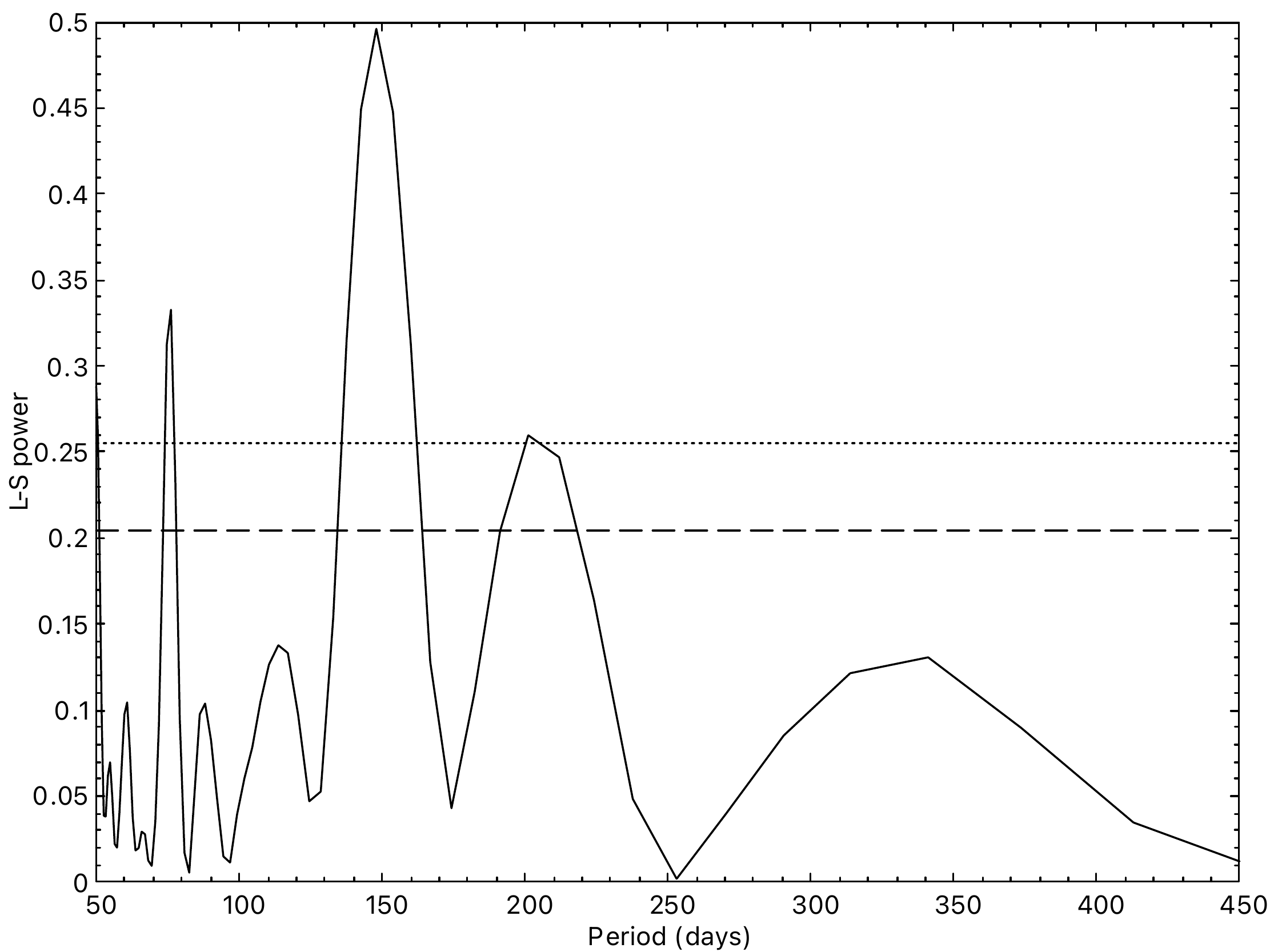}
    \caption{Lomb-Scargle analysis of the S-CUBED data. The strongest peak has a period of 148 d. The short dashed line shows the 1\% False Alarm Probability, the long dashed line the 10\% FAP.}
    \label{fig:Scubed_ls}
\end{figure}

\begin{figure*}
\resizebox{\hsize}{!}
            {\includegraphics[angle=0,width=17cm]{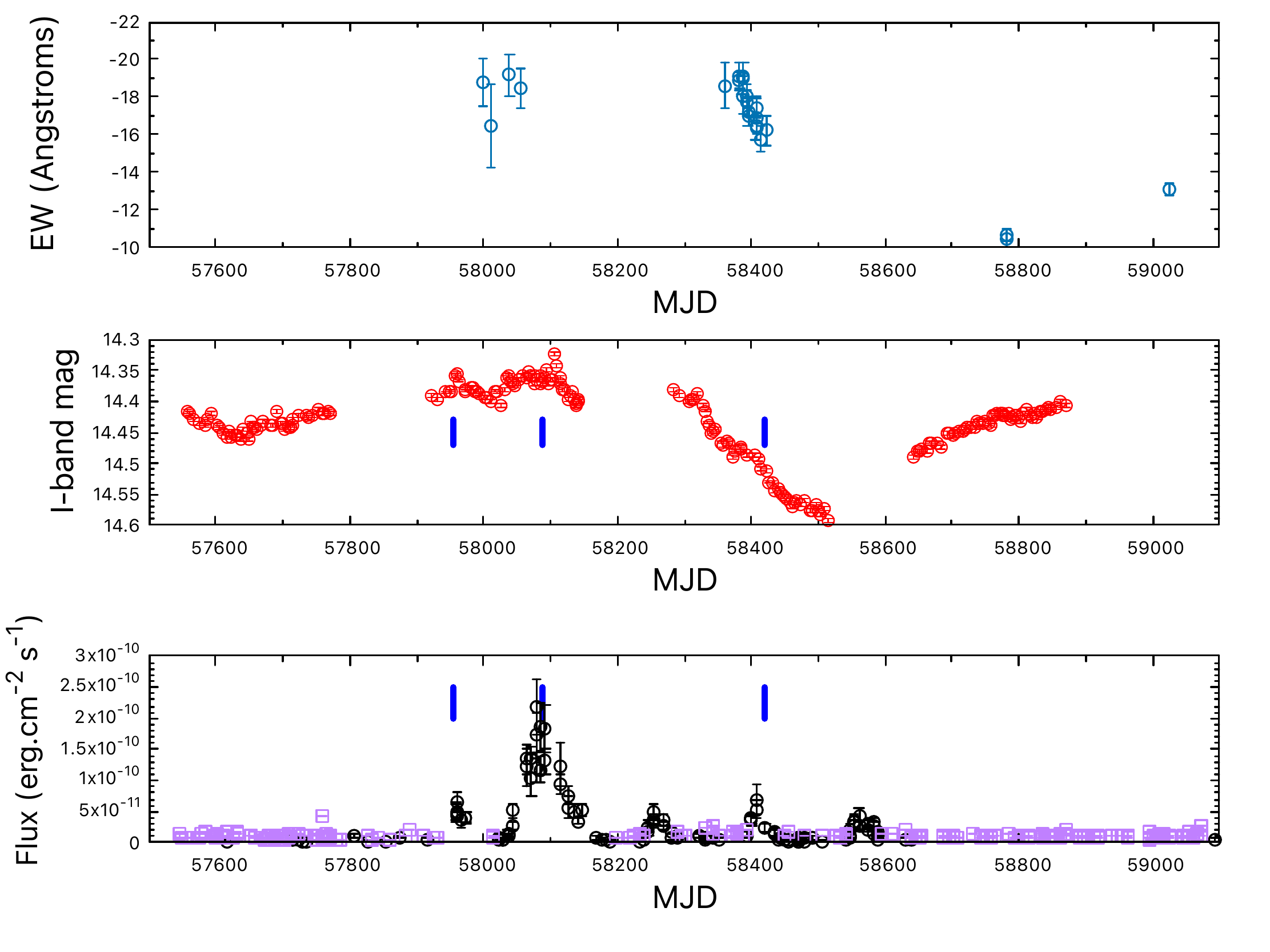}}
\caption{Long-term H$\alpha$ EW (top) and OGLE $I-$band (middle) lightcurves of SXP~15.3, together with the S-CUBED lightcurve of the combined field of SXP~15.3 and SXP~305. The black circles in the bottom plot indicate detections while the purple squares indicate upper limits. The blue markers indicate the epochs of the 15-second pulse period detections. }
\label{fig:combined_sxp15}
\end{figure*}

\begin{figure*}
            {\includegraphics[angle=0,width=14cm]{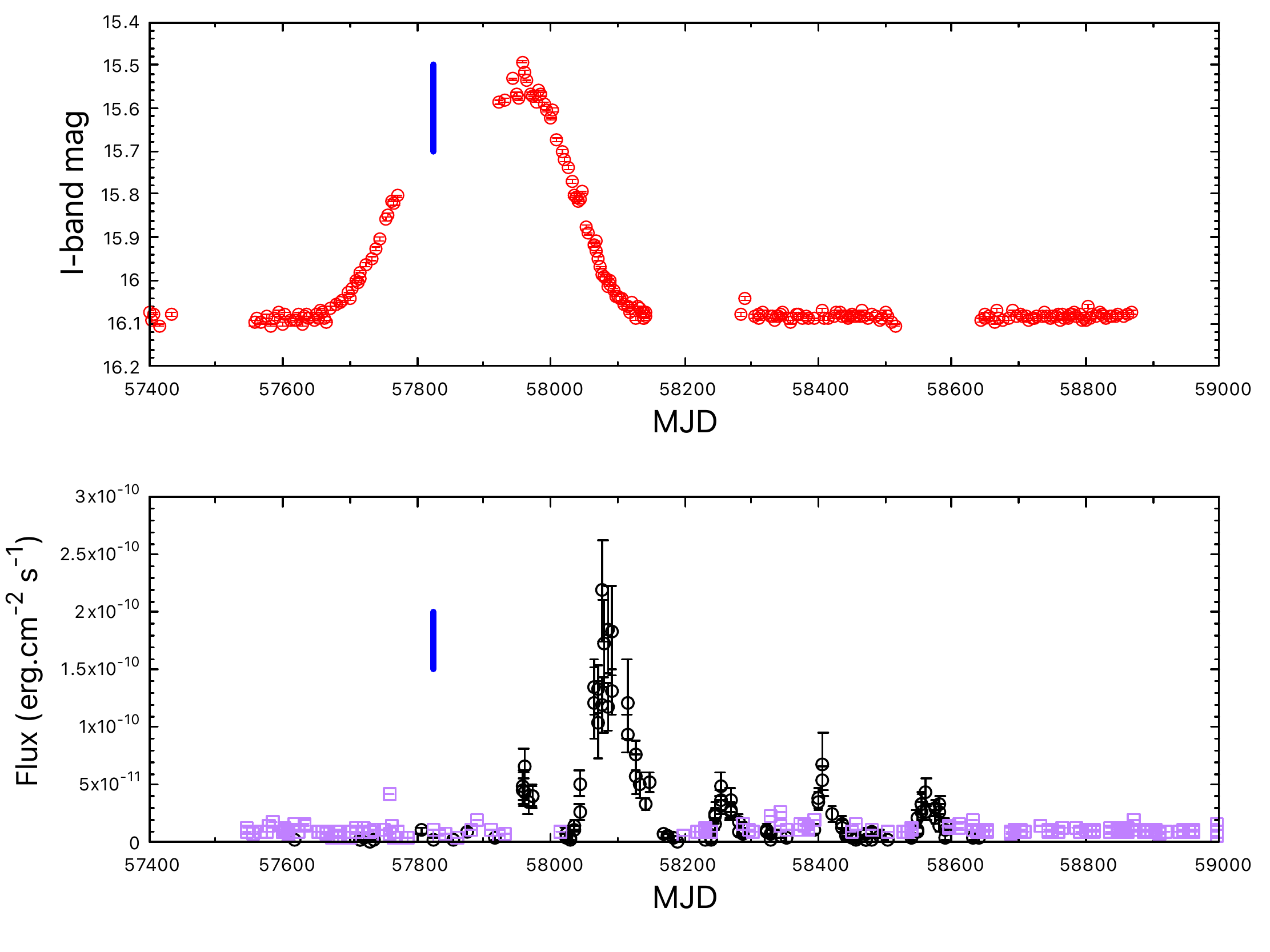}}
\caption{OGLE $I-$band lightcurve (top) of SXP~305 and the S-CUBED lightcurve of the combined field of SXP~15.3 and SXP~305. The black circles in the bottom plot indicate detections while the purple squares indicate upper limits. The blue marker indicates the epoch of the 305 s pulse period detection with NuSTAR.}
\label{fig:combined_sxp305}
\end{figure*}

The bottom panels of Figs.~\ref{fig:combined_sxp15} and \ref{fig:combined_sxp305} show the evolution of the X-ray flux from S-CUBED monitoring of the combined field of SXP~15.3 and SXP~305. At the start of the monitoring the flux displayed low values until an outburst associated with SXP~15.3 was seen in July 2017 ($\sim$MJD57960) with a peak flux of $\sim 9 \times 10^{-11}$~erg. s$^{-1}$ cm$^{-2}$. Period analysis was performed by \cite{2017ATel10600....1K} during this outburst with the 15.3~s pulse period strongly detected. Approximately four months later (November 2017) the X-ray flux of SXP~15.3 reached its highest state to ever have been detected with peak luminosity levels close to the Eddington limit ($F\sim 1.94\times 10^{-10}$~erg. s$^{-1}$ cm$^{-2}$, $L\sim 8.63\times 10^{37}$~erg. s$^{-1}$). A cyclotron feature was detected at 5~keV from \textit{AstroSAT} and \textit{NuSTAR} observations during this outburst with an estimated surface magnetic field of $6 \times 10^{11}$~G for the neutron star \citep{2018MNRAS.480L.136M}. A series of three outbursts followed, with a recurring timescale of $\sim 150$~days at peak flux levels of $\sim 4 \times 10^{-11}$~erg. s.$^{-1}$ cm$^{-2}$ ($L \sim 1.8 \times 10^{37}$~erg. s.$^{-1}$). We searched for periodic modulation of the S-CUBED data using the Generalised Lomb-Scargle technique of \cite{zk2009} where we excluded the Type II outburst seen around MJD~58100 and the results are shown in Fig.~\ref{fig:Scubed_ls}. A strong peak is seen at a period of 148~d. Interestingly, this recurring timescale is approximately double the reported orbital period of SXP~15.3 ($P = 74.5 \pm 0.05$~d; \citealt{edge2005}). We discuss this further in Section~\ref{sec:discussion}.

On delving into the RXTE archived power spectra in order to trace the history of these two sources, we found evidences of pulsations in the range of 300-310 seconds in 29 ObsIDs by employing the PUMA/ORCA pipeline\citep{laycock2005,galache2008}. The location of SXP305  lay in the central area of the PCA field of view (Collimator response>0.5) in 385 separate ObsIDs.

However, there were possible chances of some of these peaks being spurious due to aliases, spectral-leakage or due to contamination by the harmonics of other pulsars in the vicinity. On excluding such ObsIDs, there remained 10 candidate detections of SXP 305 which are listed in Table ~\ref{tab:RXTE_obs}. For one of them, an overlap was found for simultaneous detection for both SXP 15.3 and SXP 305.

For other X-ray imaging telescopes like {\it XMM-Newton}, no pulsations were found for either of the two pulsars lying in its field of view indicating towards them being in the quiescent state. Table ~\ref{tab:obs} shows the observations made by \textit{Chandra}, \nustar and \textit{Suzaku}. \nustar was the only one to successfully register strong pulsations from SXP 305 which was also reported by \cite{2019ApJ...884....2L} on 12th March 2017. The pulse profile for the same was a single broad peak. SXP 15.3 was also detected thrice by \nustar during its outburst in the period 2017-2018.

Apart from one \textit{Chandra} observation (ObsID: 7156), no significant pulsation was discovered for either of the two pulsars. However due to the excellent spatial resolution of \textit{Chandra}, it was able to resolve these closely paired X-ray binaries. So, in order to remove any ambiguity, we calculated the fluxes at the two X-ray position coordinates corresponding to the two pulsars for it. The values in the 0.5-7 keV range resembled those for quiescent BeXRB pulsars which ruled out any pulsation detection.

\begin{figure*}
            {\includegraphics[angle=0,width=12cm]{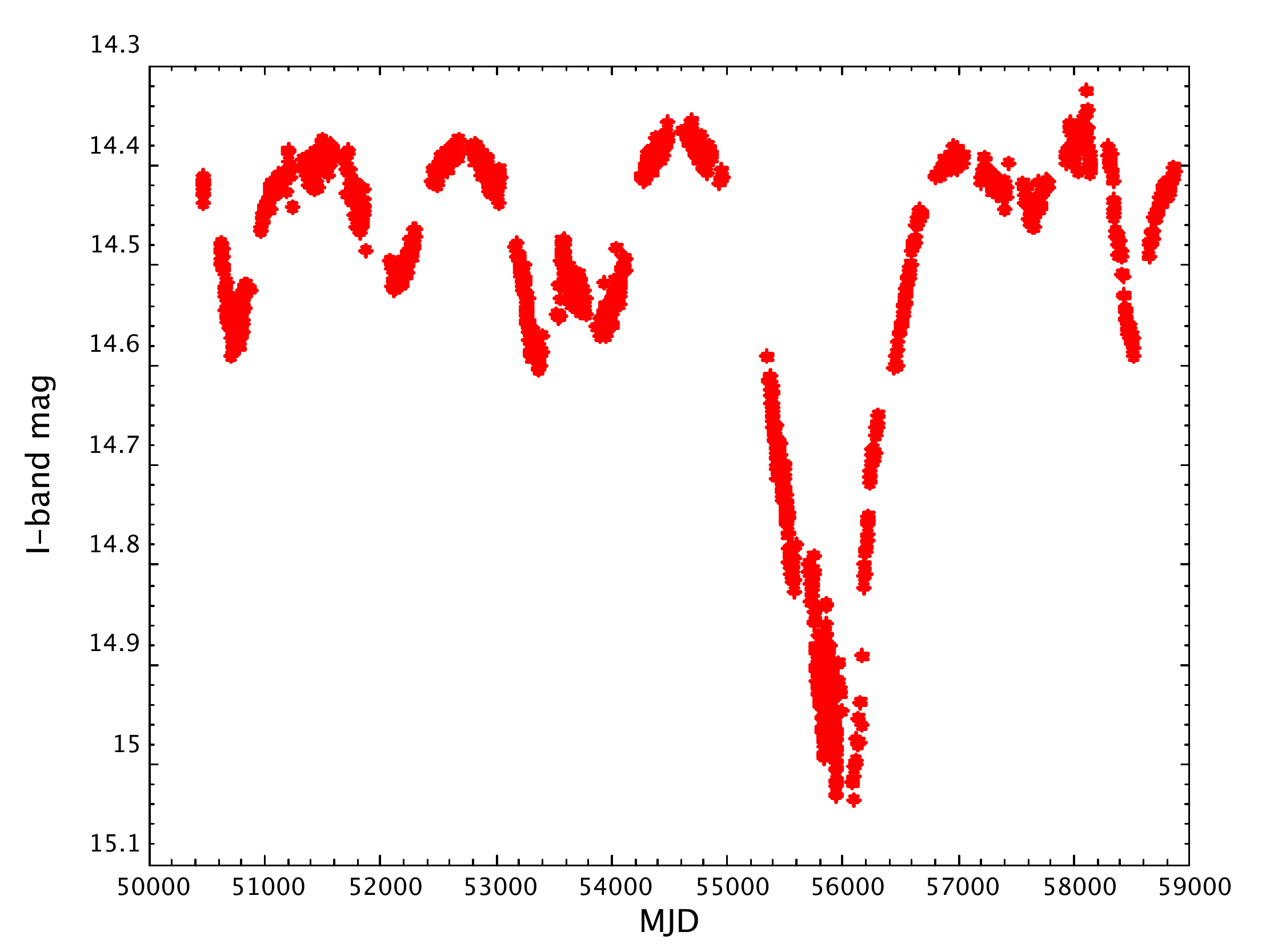}}
\caption{The full OGLE $I-$band lightcurve of the optical counterpart of SXP~15.3}
\label{fig:sxp15_full_ogle}
\end{figure*}

\begin{figure*}
            {\includegraphics[angle=0,width=14cm]{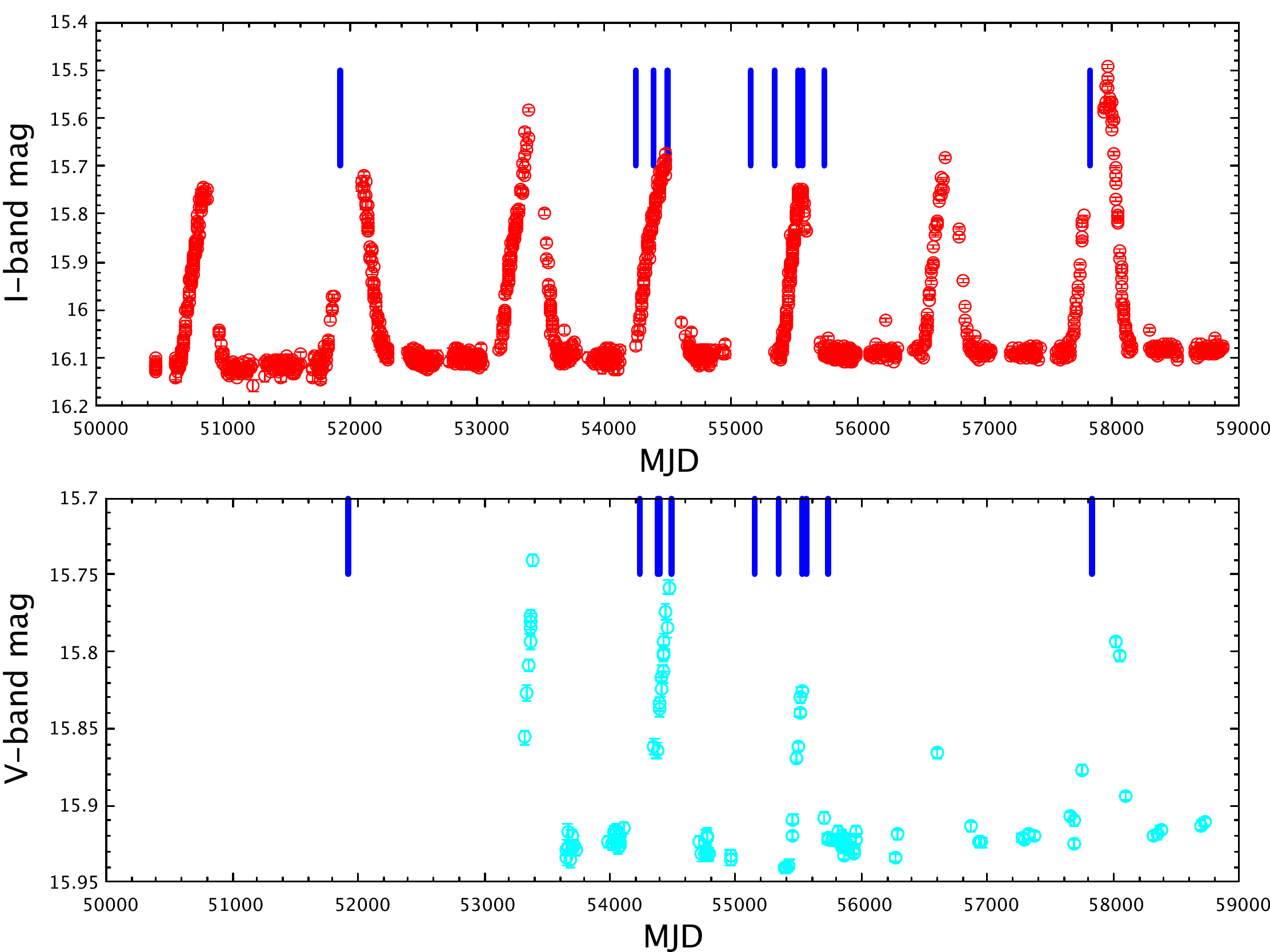}}
\caption{Long-term $I-$band (top) and $V-$band (bottom) OGLE lightcurves of the optical counterpart of SXP~305. The blue markers indicate the epochs of the 305-second pulse period detection.}
\label{fig:OGLE_VI_SXP305}
\end{figure*}

\subsection{Optical variability}
\subsubsection{SXP~15.3}
\begin{figure}
            {\includegraphics[angle=0,width=8cm]{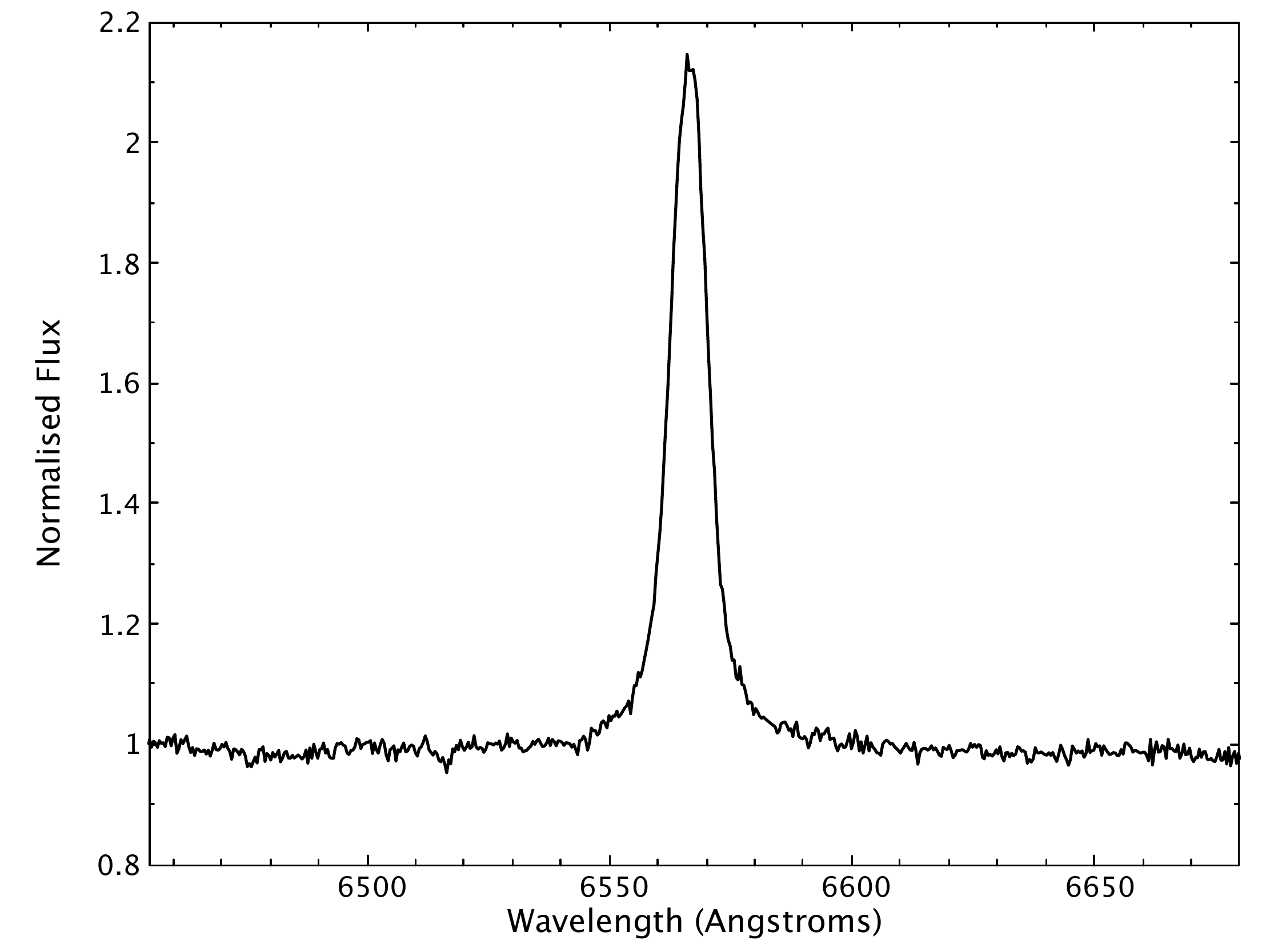}}
\caption{An example of the H$\alpha$ emission line in SXP~15.3 from an observation taken on MJD59025 (25 June 2020).}
\label{fig:sxp15_halpha}
\end{figure}

The spectroscopic monitoring of the optical counterpart of SXP~15.3 commenced during its peak photometric state when the H$\alpha$ line was seen in strong emission. Fig.~\ref{fig:sxp15_halpha} shows a typical example of the H$\alpha$ emission line. The equivalent width (EW) of the H$\alpha$ line at the start of the monitoring showed the largest values ever reported for the source, with values around $\sim -19$~\AA. The EW values continued to show correlated evolution with the OGLE flux as seen in the slow decline of the two datasets from $\sim$MJD58360.\\
The commencement of the SALT monitoring corresponds to the start of the large X-ray outburst in November 2017 ($\sim$MJD58000). This suggests that the circumstellar disc grew to a large extent, allowing enough matter to be accreted by the NS to initiate the Type II X-ray outburst. The later portion of the EW evolution ($\sim$MJD58360) shows similar high values as those seen at the start, with a gradual decrease in EW to $\sim -16$~\AA~ ($\sim$MJD58422). This period corresponds to the second of the series of outbursts in the S-CUBED lightcurve. The most recent EW measurements obtained at $\sim$MJD58785 and $\sim$MJD59025 display the lowest values ($-$10.5~\AA~and $-$13.2~\AA, respectively) of the monitoring campaign during its low X-ray state. \\

\begin{figure}
            {\includegraphics[angle=0,width=9cm]{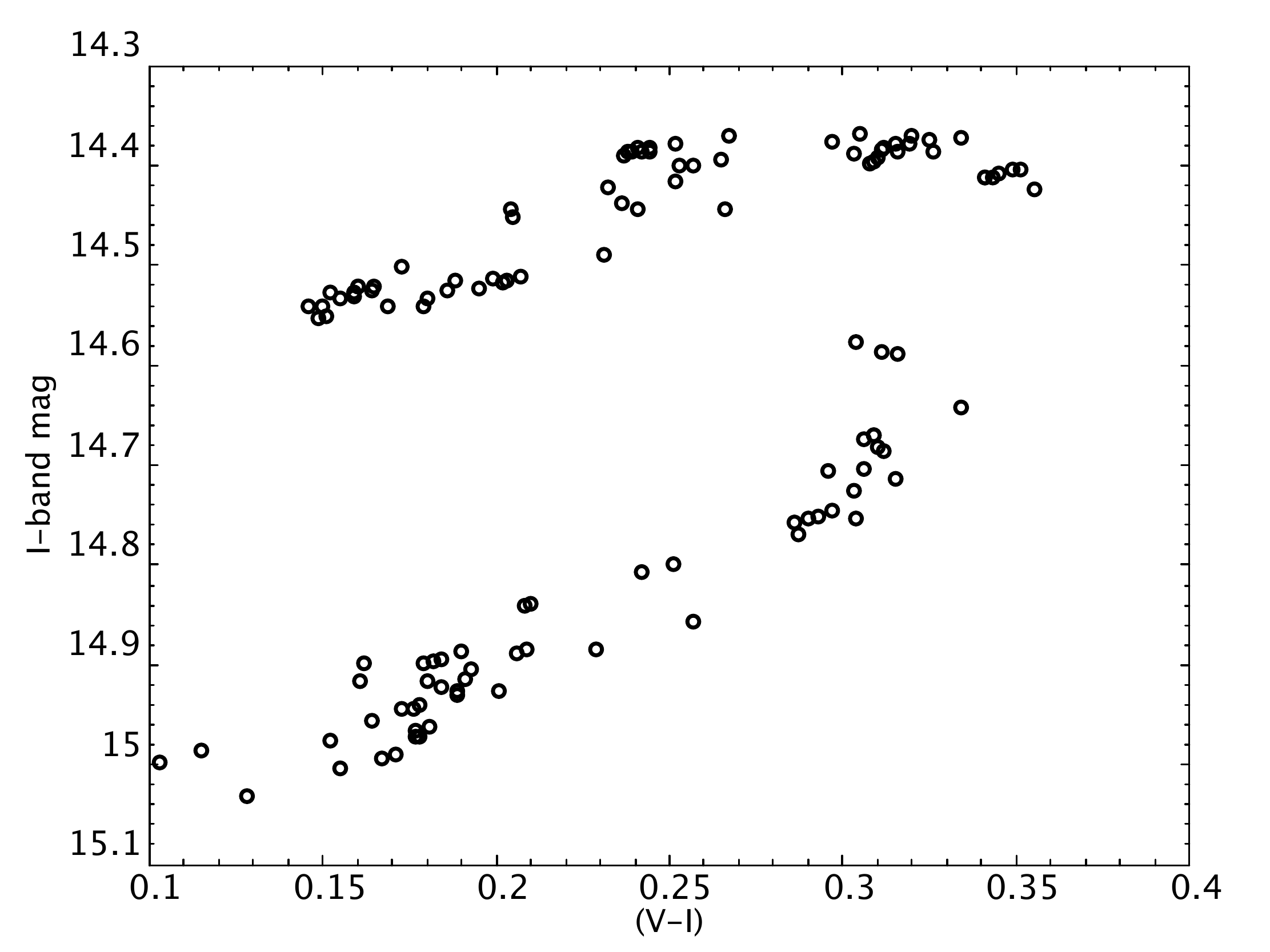}}
\caption{$(V-I)-I$ colour-magnitude plot of the optical counterpart of SXP~15.3}
\label{fig:col_mag_SXP15}
\end{figure}

The $(V-I)-I$ colour-magnitude plot for SXP~15.3 is shown in Fig.~\ref{fig:col_mag_SXP15}. A strong correlation is seen between the two quantities, which indicates a low/intermediate viewing angle of the circumstellar disc. Another notable feature of the colour-magnitude plot is a loop structure. The simultaneous $I-$ and $V-$band observations were obtained during the period between MJD53000 and MJD59000, covering the time when the system underwent its lowest dip in flux (Fig.~\ref{fig:sxp15_full_ogle}). From Fig.~\ref{fig:col_mag_SXP15} it appears that the disc is lost from the inside out since there was no evidence of the enhanced accretion (i.e. an outburst) during that time that could have disrupted the outer parts of the disc.

\begin{figure}
    \centering
    \includegraphics[width=9cm]{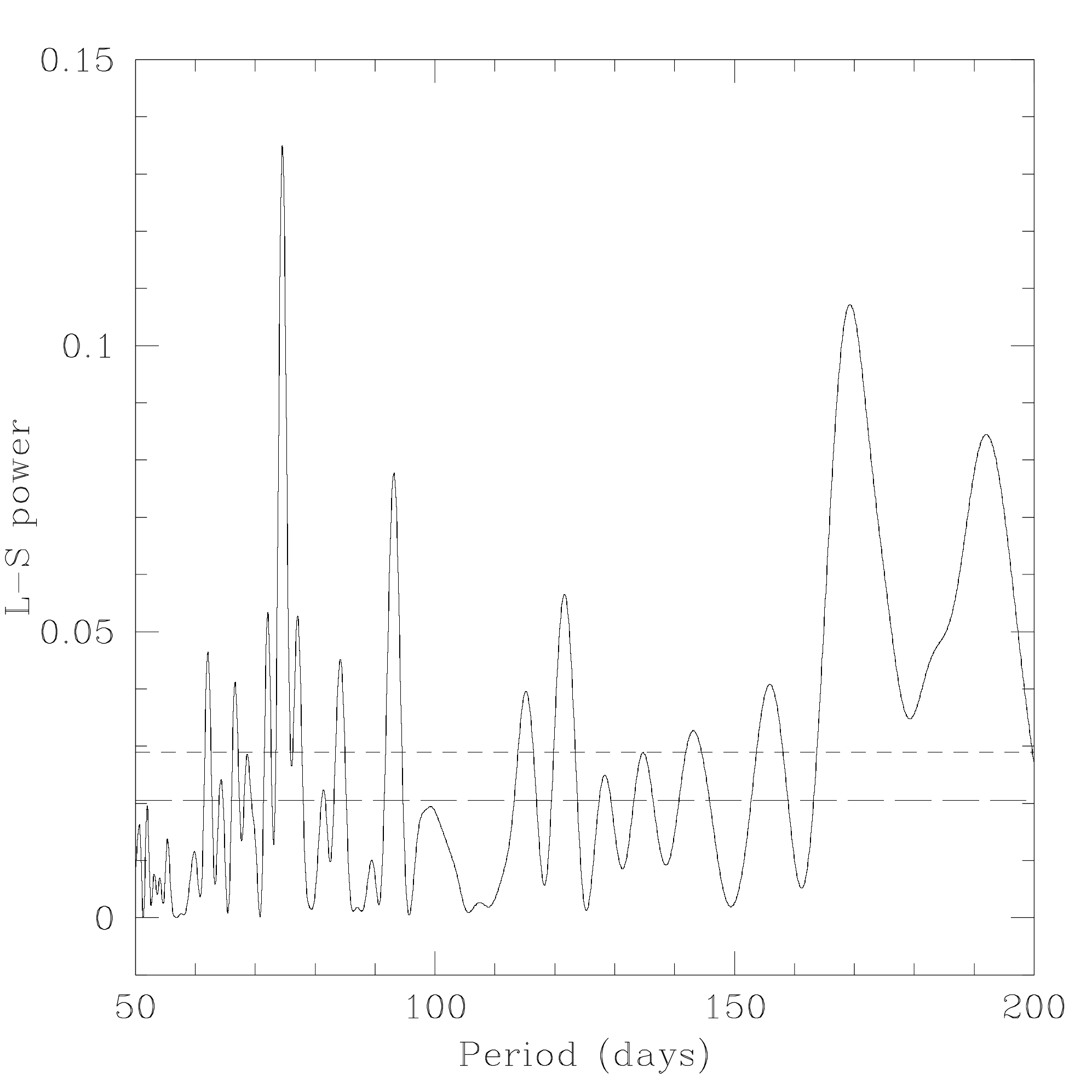}
    \caption{Lomb-Scargle analysis of OGLE IV I-band data. The strongest peak has a period of 74.5 d. The short dashed line shows the 1\% False Alarm Probability, the long dashed line the 10\% FAP. The peak at 94 d is the beat period between the 74.5 d period and the annual sampling.}
    \label{fig:o4ls}
\end{figure}

If the OGLE IV data are searched for periodic modulation in the range 50 -- 200 d using the Generalised Lomb-Scargle then the result is shown in Fig ~\ref{fig:o4ls}. This confirms the same period that has been identified by previous authors \citep{edge2005, 2011MNRAS.413.1600R} in the the earlier MACHO and OGLE III data sets - 74.5 d. The precise updated ephemeris for the time of the optical peaks, $T_{opt}$, is:

\begin{equation}
T_{opt} = 2452085.9 + N(74.516) ~\textrm{JD}\label{eq:1}
\end{equation}

\subsubsection{SXP~305}
\begin{figure}
\resizebox{\hsize}{!}
            {\includegraphics[angle=0,width=17cm]{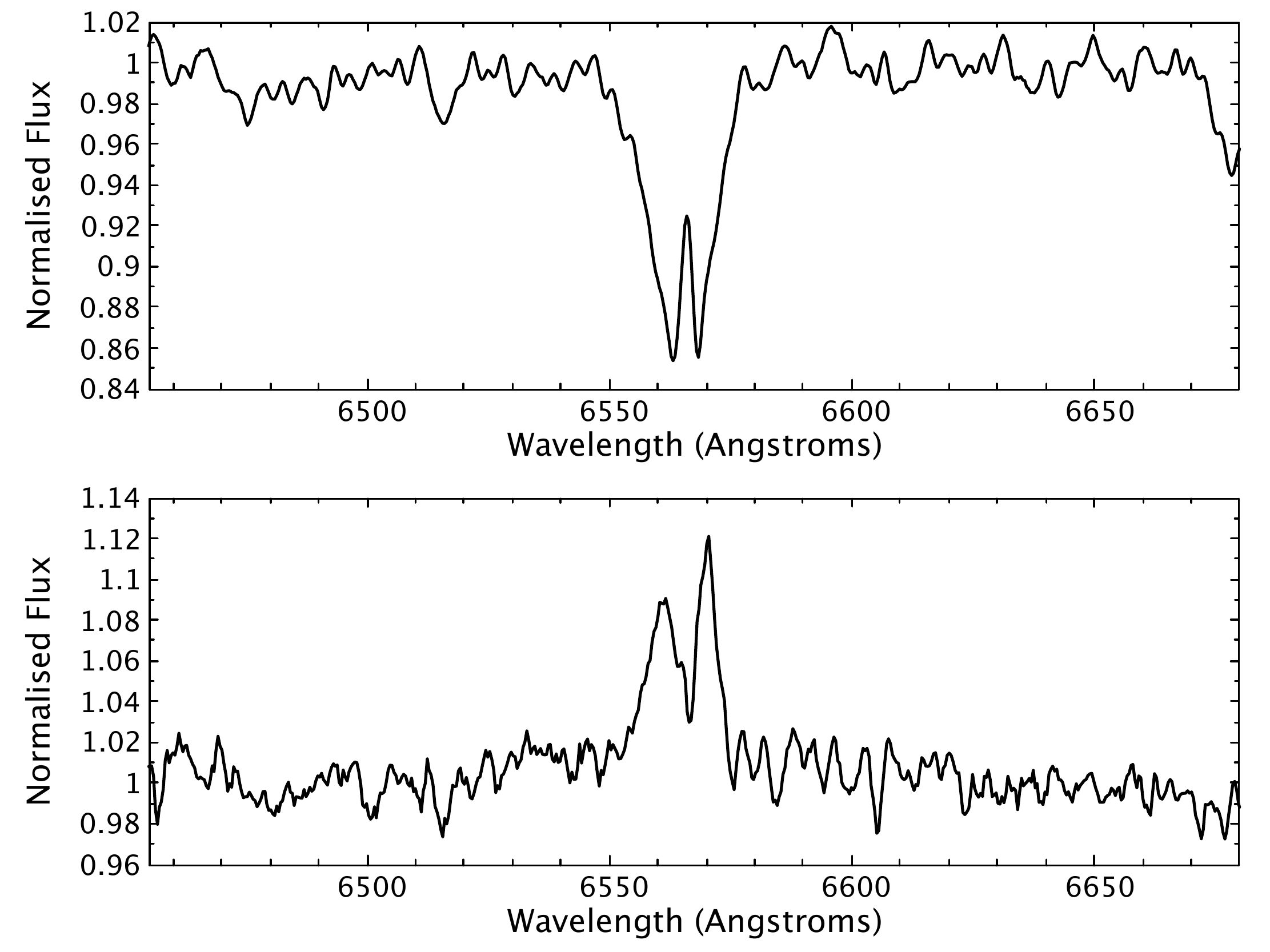}}
\caption{H$\alpha$ line of the optical counterpart of SXP~305 from SALT. The top panel profile is from the first observation obtained on MJD58785 (29 October 2019) and the bottom panel is a profile obtained on MJD59025 (25 June 2020).}
\label{fig:SXP305_Halpha}
\end{figure}

\begin{figure}
            {\includegraphics[angle=0,width=9cm]{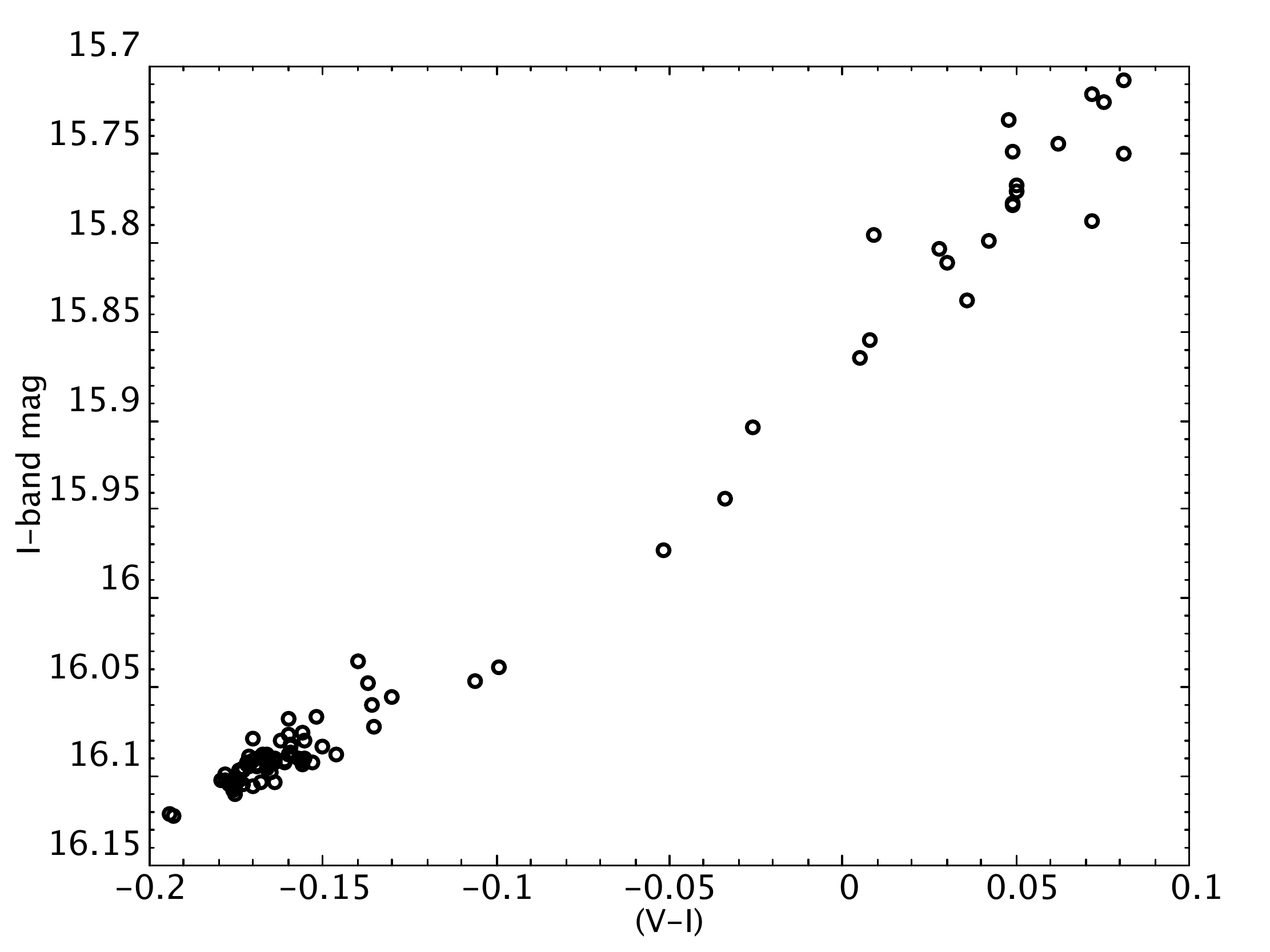}}
\caption{$(V-I)-I$ colour-magnitude plot of the optical counterpart of SXP~305}
\label{fig:col_mag_SXP305}
\end{figure}

The OGLE lightcurve shows repeating outbursts with I-band amplitudes in the range $\sim 0.35-0.6$~mag that last for up to $\sim$600~days (Fig.~\ref{fig:OGLE_VI_SXP305}). The recurring timescale of these outbursts has a range $1100-1400$~days. \\
The first SALT observation of the optical counterpart of SXP~305 covering the red region of the spectrum shows the H$\alpha$ line in absorption (Fig.~\ref{fig:SXP305_Halpha}, top panel). This observation was taken during its OGLE photometric minimum ($\sim$MJD58785). In the most recent observation (MJD59025) the H$\alpha$ line is seen in emission for the first time ever (Fig.~\ref{fig:SXP305_Halpha}, bottom panel), confirming the Be nature of the optical companion. OGLE was unavailable during this time for monitoring, however, the emission line suggests that SXP~305 is undergoing another optical outburst. This variability is an indicator of extreme changes in the circumstellar disc, suggesting that the system goes through regular periods of complete disc-loss and recovery (see section~\ref{sec:origin}). \\
Fig.~\ref{fig:col_mag_SXP305} shows the $(V-I)-I$ colour-magnitude plot of the optical counterpart of SXP~305. The relationship between the two quantities shows a strong correlation, indicating a low/intermediate viewing angle of the circumstellar disc. Growth in radius from a circumstellar disc that has a low inclination angle relative to our line of sight results in excess flux and reddening due to the temperature of the outer disc regions being cooler than the star.

\subsubsection{Spectral classification}
\label{sec:spec_class}
\begin{figure*}
            {\includegraphics[angle=0,width=13cm]{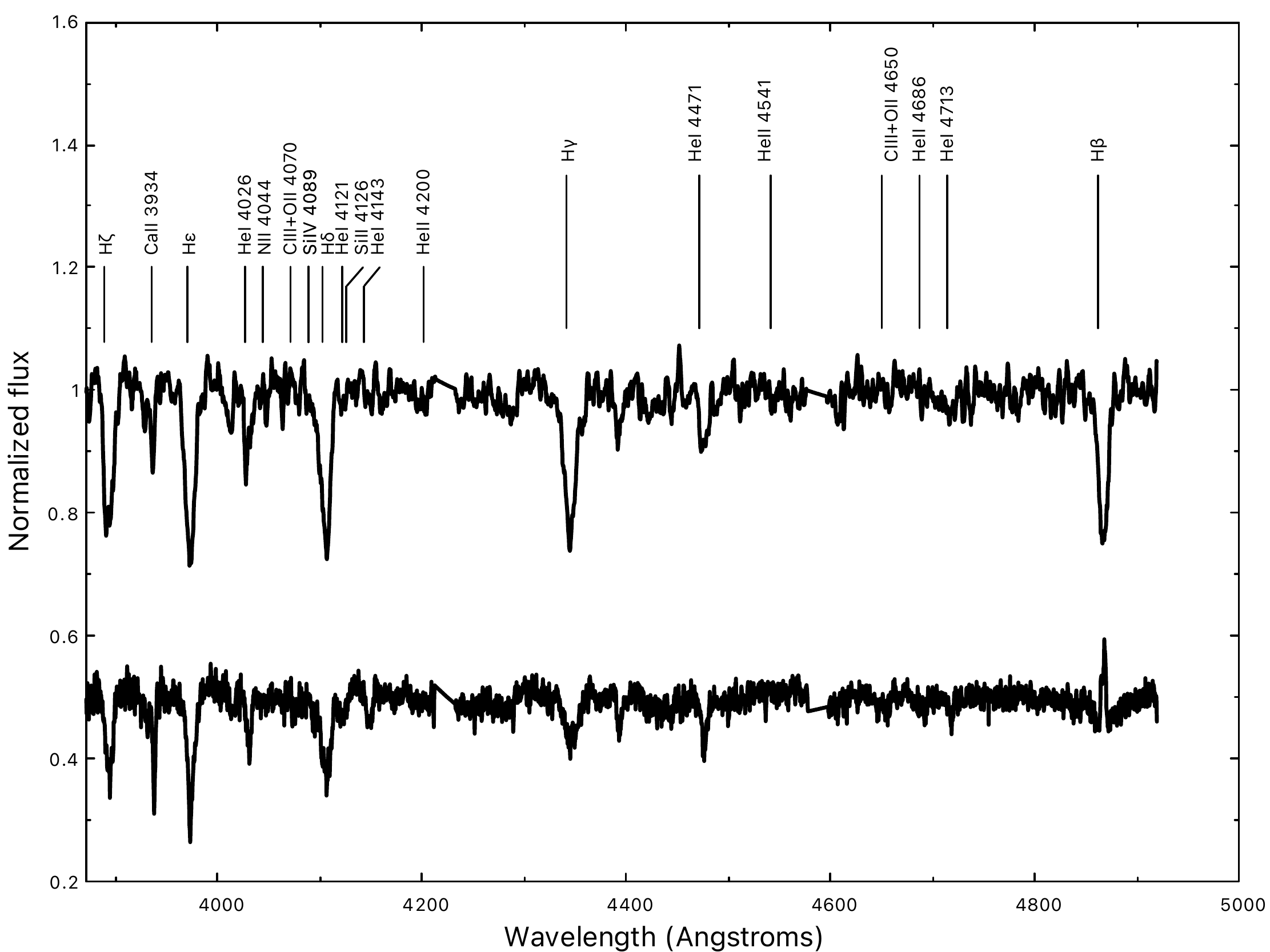}}
\caption{SALT spectra of SXP~305 (top) and SXP~15.3 (bottom) covering the blue region. The spectra are corrected for the heliocenter and redshift of the SMC. The different line species are labeled at their expected rest wavelengths.}
\label{fig:blue_spec}
\end{figure*}

The blue spectra of the optical counterparts of SXP~15.3 and SXP~305 are shown in Fig.~\ref{fig:blue_spec}. Both spectra are clearly those of early-type stars, with a number of Balmer and helium lines present in absorption. The H$\beta$ line in the spectrum of SXP~15.3 is seen in emission due to the presence of the circumstellar disc. A few of the absorption lines show infilling as a result of the emission from the circumstellar disc. \\
Using the criteria in \cite{2004MNRAS.353..601E}, the presence of the CIII and OII blend constrains the spectral type to earlier than B3. The presence of the weak HeII 4686 line and absence of the He II 4541 line further constrains the spectral type to B0.5. Using the faintest observed V-band magnitude of 15.15 and distance modulus of the SMC of 18.95 \citep{2013IAUS..289..222G}, we constrain the luminosity class to V \citep{1981Ap&SS..80..353S, 2013ApJS..208....9P}. We conclude that the spectral type of SXP~15.3 is therefore B0.5 V, in agreement, to within errors, with \cite{2001A&A...374.1009C} and \cite{2008MNRAS.388.1198M}.\\

Using the same criteria, the presence of the CIII and OII blend at $\sim$4650~\AA\ constrains the optical counterpart of SXP~305 to B3 or earlier while the absence of the HeII~4686 line sets the spectral type to later than B0.5. The spectrum reveals the SiIV 4088 line being weaker than the OII 4415-17 blend, which constrains the spectral type to B1.5 or earlier. The low signal-to-noise of the spectrum makes it difficult to confirm the presence/absence of the other weaker indicators from the noise and so we conclude that the spectral type of this object is in the range B0.5$-$B1.5. Using the faintest observed V-band magnitude of 15.95 constrains the luminosity class to V \citep{1981Ap&SS..80..353S, 2013ApJS..208....9P}. In conclusion, the spectral type of the optical counterpart of SXP~305 is B0.5$-$B1.5 V.

\section{Discussion}
\label{sec:discussion}
\subsection{BeXB nature of SXP~305}
\cite{2019ApJ...884....2L} confirmed the 305-sec pulse period and concluded that SXP~305 is a BeXB system, although the Be nature of the optical companion had not been determined. The presence of the H$\alpha$ emission line and the B-type spectral classification presented in this work from SALT observations confirms the Be nature of the massive companion. The previous observation covering the red wavelength region was obtained during the optical minimum as seen in the OGLE lightcurve when the disc was absent \citep{2000A&A...354..999K}. \\

\subsection{Origin of the recent series of Type I X-ray outbursts}
\label{sec:origin}
A disc-less B1 V spectral class star would have an apparent I-band magnitude of $\sim 16.3$ and $(V-I)$ colour index of $\sim-0.19$ \citep{1981Ap&SS..80..353S, 2013ApJS..208....9P}. These values lie at the bottom left-hand corner of Fig.~\ref{fig:col_mag_SXP305}. This implies that at its faintest state, the Be disc in SXP~305 is almost completely diminished. This is further supported by the absence of the H$\alpha$ emission line during the OGLE minima (Fig.~\ref{fig:SXP305_Halpha}). The shape and repetition of the optical outbursts suggests quasi-periodic mass-loss from the Be star. From an examination of the S-CUBED vs. I-band plots for both systems (Figs.~\ref{fig:combined_sxp15} and \ref{fig:combined_sxp305}) it looks unlikely that the X-ray emission from the recent outbursts is dominated by SXP~305 since this is the time of photometric minimum, when the disc was depleted.  \\
This recent series of X-ray outbursts occurred during the decline of the optical emission (both photometric and EW) of SXP~15.3, with the spectroscopic observations covering the second of the three outbursts (Fig.~\ref{fig:combined_sxp15}). Therefore, it is likely that these outbursts come from SXP~15.3. \\
The pulse period detections of SXP~15.3 from NuSTAR were all obtained close to outburst peaks, with the last of them occurring during the the second of the three recent recurring outbursts (Fig.~\ref{fig:combined_sxp15}). The NuSTAR detection of the 305~second pulse period in SXP~305 corresponds to an epoch when there was no enhanced X-ray activity which is also consistent with the time when the I-band emission was on the rise (Fig.~\ref{fig:combined_sxp305}). This possibly suggests that there was some low-level accretion from the Be disc in SXP~305 as it was forming during this time. The fact that a strong pulse period was detected from SXP~15.3 during one of the recent outbursts is further evidence that the outbursts are associated with SXP~15.3 instead of SXP~305.

\subsection{Timing analysis of the recent series of Type I X-ray outbursts}
\subsubsection{The true orbital period of SXP~15.3}
\begin{figure}
            {\includegraphics[angle=0,width=7.5cm]{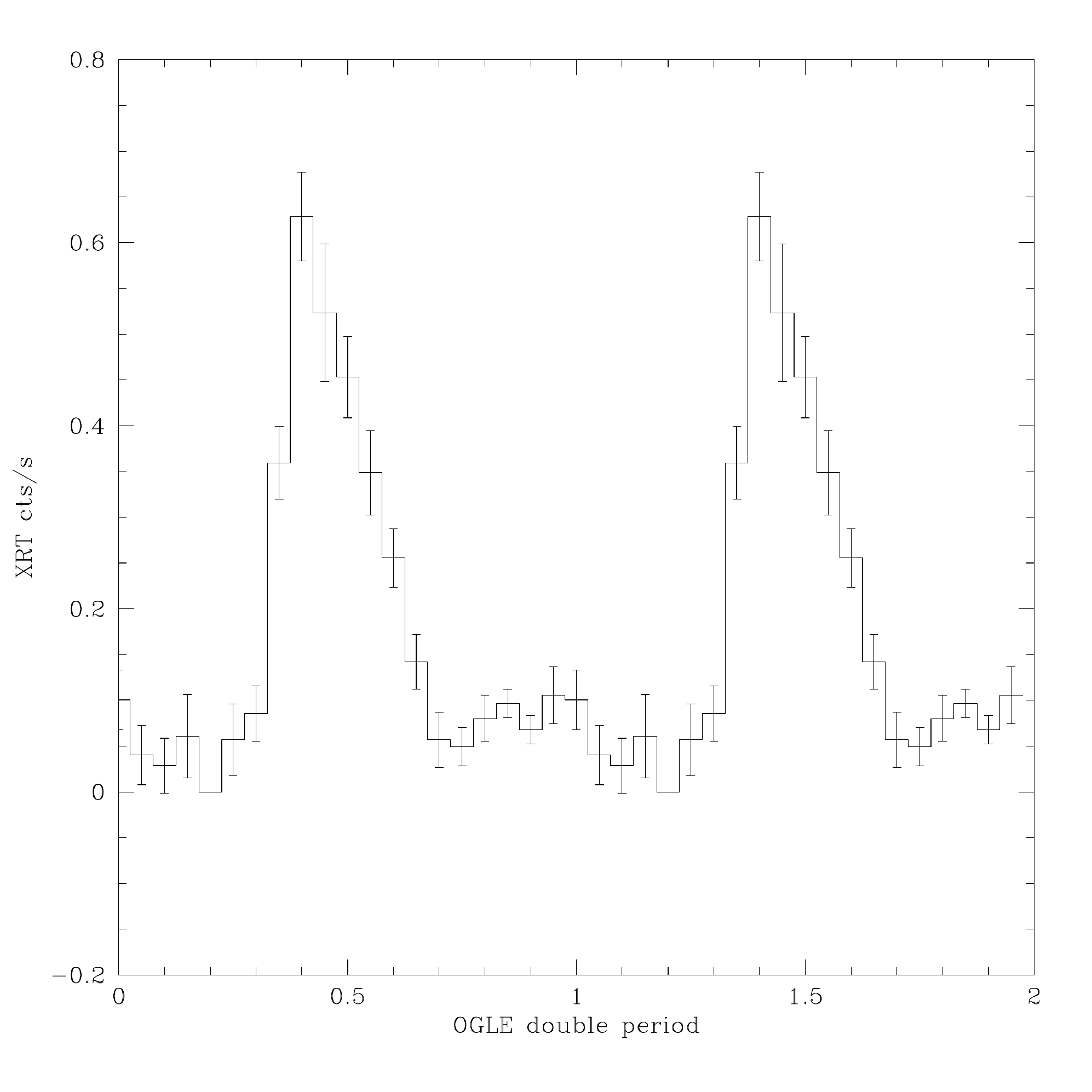}}
\caption{S-CUBED data folded on the 148~d period obtained in Fig.~\ref{fig:Scubed_ls}.}
\label{fig:sfold}
\end{figure}

\begin{figure}
            {\includegraphics[angle=0,width=7.5cm]{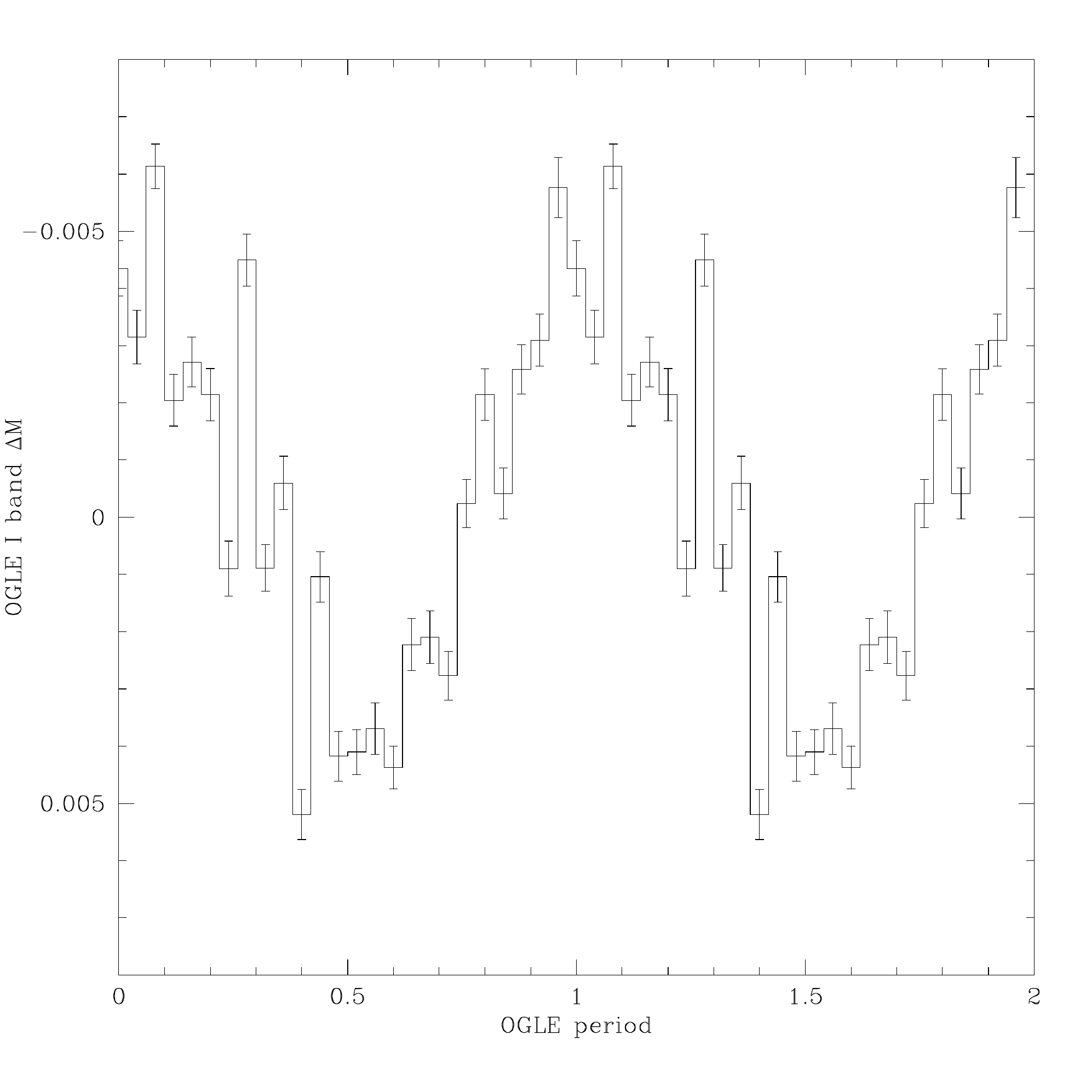}}
\caption{OGLE detrended data folded at the OGLE period given in Eqn ~\ref{eq:1}.}
\label{fig:fold_period}
\end{figure}

If data from the Type II outburst seen around MJD 58100 are removed from the S-CUBED light curve, and the remaining data are folded at the period obtained from the Lomb-Scargle analysis (Fig.~\ref{fig:Scubed_ls}), then the resulting profile clearly indicates the presence of a very significant X-ray peak every 148.0 d - see Fig.~\ref{fig:sfold}. 
Fig.~\ref{fig:fold_period} shows a similar version of this plot, but with the optical data folded on the 74.5-day period.

It is clear from the timing analysis of the optical data and the recurrence time of the X-ray outbursts in SXP~15.3 that there are two periodicities of 74.5 days and 148 days, respectively. Here, we explore the possible physical interpretations of the recent X-ray outbursting behaviour, where we consider either of these two periodicities being the true orbital period of the system. Both the optical and X-ray peaks extend over a large fraction of the orbit for both period considerations. While the optical peak is a proxy of the Be disc distortion, the X-ray peak is a measure of the accretion of matter onto the neutron star. The long time span of these peaks could be an indicator of a relatively small misalignment between the Be disc and neutron star orbital planes, as well as a relatively low eccentricity of the neutron star orbit. This would imply that the neutron star is in the neighbourhood of the Be disc and distorts it, while accreting matter for a significant portion of its orbit.\\
For the case where the 74.5-day periodicity is the orbital period, this would mean that the OGLE $I-$band outburst occurs once per orbit while the X-ray outburst profile would occur at every other orbit of the neutron star (Fig.~\ref{fig:fold_period}). In a geometrical setup where the Be disc and the orbital plane are coplanar, with the neutron star in a nearly circular orbit, it is possible for the disc eccentricity to grow through the 3:1 resonance \citep{2019ApJ...881L..32F}. If the disc precesses apsidally in a prograde direction, the time interval of the X-ray outbursts can be greater than the orbital period.\\
In the instance where the 148-day periodicity is the true orbital period, this would imply that the optical outbursts occur twice per orbit while the X-ray outburst profile shows a peak once per orbit (Fig.~\ref{fig:sfold}). In this scenario, it is possible for the orientation of the Be disc and neutron star orbital planes to be close enough for the disc to be distorted for a large portion of the orbit, but with enough misalignment to generate two optical outbursts per orbit. \\
A further piece of evidence in support of the longer 148 d orbit comes from interpreting the H$\alpha$ EW value as a circumstellar disc size. Using the method described in \cite{hanu1989} and the measured value of -19\AA ~the disc radius is estimated to be $9.6\times10^{10}$ m. Kepler's Laws predict that the radius of a circular orbit with these stellar components (1.4$M_\odot$ and 17.5$M_\odot$) and a period of 14 d will be $9.3\times10^{10}$ m, whereas a 75 d orbit size would only be $5.8\times10^{10}$ m. Since it is believed that the neutron star orbit naturally curtails the circumstellar disk growth to be inside its orbit \citep{okazaki2001}, then a 75 d orbit would not work well under that paradigm. A 148 d orbit fits much better.

\subsubsection{Optical and X-ray peak delays}
As Fig.~\ref{fig:sfold} shows, the X-ray peak tends to precede the optical one. There are 4 recognisable Type I X-ray peaks in the S-CUBED light curve, apart from the large Type II outburst. These occurred at dates (OGLE phase): MJD 57962 (0.86), MJD 58259 (0.84), MJD 58410 (0.87) and MJD 58565 (0.95). So when the regular Type I X-ray outbursts occur they are always at a phase 0.15 - 0.05 before the peak optical brightness. This type of phase delay has been seen in several other systems \citep{coe2015} with values in the range 0.04 - 0.10, corresponding to actual delays of 2 - 13 d. In the case of SXP~15.3, the delay of the optical peak with respect to the X-ray peak of $0.1 \pm 0.05$ corresponds to period of $7 \pm 4 d$, consistent with previous values. It is suggested in \cite{coe2015} that this delay is characteristic of the circumstellar disk response times within the accretion process, and it is the timing of the X-ray outburst that marks the periastron passage of the neutron star. Unfortunately a lack of pulse period measurements around the orbit for SXP~15.3 prevent an accurate orbital solution being obtained to confirm either the exact moment of periastron or the true orbital period.

\section{Conclusions}

We have presented long-term optical and X-ray observations of two neighbouring SMC Be X-ray binaries, SXP~15.3 and SXP~305. The recent X-ray data reveal periodic outbursting behaviour from the combined region of the two objects, the origin of which we demonstrate is likely due to SXP~15.3. We confirm the previously reported 74-day modulation from the optical data of SXP~15.3, which implies that the optical outbursts occur at twice the frequency of the X-ray outbursts. We have considered either of the optical and X-ray modulation as the true orbital period of SXP~15.3 and propose physical scenarios to explain the observational behaviour. In both cases we suggest that the modulation is caused by an unusual disc orientation as a result of a complex configuration between the Be disc and the orbital plane. Further analysis in the form of hydrodynamic modelling, or pulse period measurements around the orbit, are necessary to distinguish between the two possible orbital periods.

\section*{Acknowledgements}

We thank an anonymous referee for the useful comments. IMM, DAHB and LJT are supported by the South African NRF. Some of the observations reported in this paper were obtained with the Southern
African Large Telescope (SALT), as part of the Large Science Programme on transients 2018-2-LSP-001 (PI: Buckley). This work was partially supported by the NASA ADAP grants NNX14AF77G and 80NSSC18K0430. 

\section*{Data Availability}

 All X-ray data are freely available from the NASA Swift archive. The OGLE data in this article will be shared on any reasonable request to Andrzej Udalski of the OGLE project. The SALT data in this article will be shared on any reasonable request to David Buckley of the South African Astronomical Observatory.
 


\bibliographystyle{mnras}
\bibliography{references} 



\appendix

\section{X-ray data}


\begin{table*}
\begin{small}
	\begin{centering}
	\caption{}
	\label{tab:obs}
	\begin{tabular}{cccccccccc}
		\hline
		\textbf{Obs. ID} & \textbf{Time} &\textbf{MJD}  & \textbf{RA} & \textbf{Dec}  & \textbf{Flux}/$10^{-14}$ & \textbf{L\textsubscript{x}}/$10^{34}$ &  \textbf{Period} \\
		& (ks) & (d) & (deg)  & (deg) & (erg~s$^{-1}$~cm$^{-2}$)  &(erg~s$^{-1}$) & (s) \\
		\hline
		\textbf{Chandra} & \textbf{ACIS-I} & & & &\textbf(0.5-7 keV)& & &  \\
		\hline
		7156 & 38.67 & 54061 & 13.0583 & -73.3219 & 2.44$\pm$0.57 & 1.12$\pm$0.26 & 15.239  \\
		     &      &       & 13.0644          &  -73.3209      &      5.52$\pm$0.81         &       2.53$\pm$0.37        & ...        \\
		8479 & 42.10 & 54060 & 13.0583 & -73.3219 & 3.82$\pm$0.66 & 1.75$\pm$0.30 & ...  \\
		 &      &       & 13.0644          &  -73.3209        &    4.73$\pm$0.72 &  2.18$\pm$0.33             & ...       \\
		8481 & 15.98 & 54062 & 13.0583 & -73.3219 & 3.3$\pm$1.1 & 1.52$\pm$0.50 & ... \\
		 &      &       & 13.0644          &  -73.3209         &      4.71$\pm$1.18         &  2.16$\pm$0.54             & ...         \\
		\hline
		\textbf{Chandra} & \textbf{ACIS-S} & & & &\textbf(0.5-7 keV)&&& \\
		\hline
		3907  & 50.19 & 52698 & 13.0583 & -73.3219 & 2.68$\pm$0.55 & 1.23$\pm$0.25 & ...  \\
		&      &       & 13.0644          &  -73.3209        &      5.90$\pm$0.77         &  2.71$\pm$0.35             & .....     \\
			11095 & 38.67 & 55334 & 13.0583 & -73.3219 & ... & ... & ... \\
		     &      &       & 13.0644          &  -73.3209       &        ...       &       ...        & ...     \\
			11096 & 38.67 & 55349 & 13.0583 & -73.3219 & ... & ... & ... \\
		     &      &       & 13.0644          &  -73.3209        &          ...     &       ...        & ...        \\
		11097 & 29.60 & 55300 & 13.0583 & -73.3219 & 0.33$\pm$0.32 & 0.15$\pm$0.15 & ...  \\
			&      &       & 13.0644          &  -73.3209         &    1.97$\pm$0.48           & 0.91$\pm$0.22              & ....        \\
		11980 & 22.75 & 55307 & 13.0583 & -73.3219 & 0.65$\pm$0.31 & 0.30$\pm$0.14 & ...  \\
			&      &       & 13.0644          &  -73.3209          &       1.76$\pm$0.51        &  0.81$\pm$0.23            & .....       \\
		11981 & 33.60 & 55317 & 13.0583 & -73.3219 & 1.01$\pm$0.31 & 0.46$\pm$0.14 & ...  \\
		&      &       & 13.0644          &  -73.3209         &    1.69$\pm$0.42           &   0.77$\pm$0.19            & ....      \\
		11982 & 24.57 & 55319 & 13.0583 & -73.3219 & 1.36$\pm$0.78 & 0.62$\pm$0.36 & ... \\
		&      &       & 13.0644          &  -73.3209         &    1.01$\pm$0.76           &    0.46$\pm$0.35           & ....      \\
			11983 & 29.34 & 55328 & 13.0583 & -73.3219 & ... & ... & ... & - \\
		     &      &       & 13.0644          &  -73.3209         &          ...     &       ...        & ...   \\
			11984 & 20.48 & 55338 & 13.0583 & -73.3219 & ... & ... & ... & - \\
		     &      &       & 13.0644          &  -73.3209         &          ...     &       ...        & ...    \\
		 11985 & 21.04 & 55341 & 13.0583 & -73.3219 & ... & ... & ... & - \\
		     &      &       & 13.0644          &  -73.3209         &          ...     &       ...        & ...    \\
		12200 & 26.73 & 55308 & 13.0583 & -73.3219 & 0.55$\pm$0.28 & 0.25$\pm$0.13 & ... \\
		 &      &       & 13.0644          &  -73.3209         &          1.91$\pm$0.51     &       0.88$\pm$0.23        & ...        \\
		12208 & 15.97 & 55318 & 13.0644          &  -73.3209 & 1.22$\pm$0.52 & 0.56$\pm$0.24 & ...  \\
		12210 & 27.55 & 55323 & 13.0583 & -73.3219 & 0.68$\pm$0.43 & 0.31$\pm$0.20 & ... \\ 
		&      &       & 13.0644          &  -73.3209         & 1.06$\pm$0.87              & 0.49$\pm$0.40             & .....       \\
		12211 & 34.99 & 55320 & 13.0583 & -73.3219 & 1.13$\pm$0.44 & 0.52$\pm$0.20 & ... \\ 
		&      &       & 13.0644          &  -73.3209         &    0.95$\pm$0.57           &      0.44$\pm$0.26        & .....        \\
 		\hline
		\textbf{XMM}   & \textbf{EPIC} & & & &\textbf(0.2-12 keV)&&& \\
		\hline
		0110000101 & 28.00 & 51832 & 13.0639 & -73.3204 &
		\begin{tabular}[c]{@{}c@{}}3.70$\pm$2.44\end{tabular} &\begin{tabular}[c]{@{}c@{}}1.70$\pm$1.12\end{tabular}& ... & \begin{tabular}[c]{@{}c@{}}\end{tabular} \\ 
		
		0404680301 & 23.90 & 54201 & 13.0642 & -73.3203 &
		\begin{tabular}[c]{@{}c@{}}8.12$\pm$1.51\end{tabular} &\begin{tabular}[c]{@{}c@{}}3.73$\pm$0.69\end{tabular}& ... & \begin{tabular}[c]{@{}c@{}}\end{tabular} \\ 
		
		0601213001 & 42.01 & 55087 & 13.0617 & -73.3212 &
		\begin{tabular}[c]{@{}c@{}}13.93$\pm$3.74\end{tabular} &\begin{tabular}[c]{@{}c@{}}6.41$\pm$1.72\end{tabular}& ... & \begin{tabular}[c]{@{}c@{}}\end{tabular}\\ 
		
		0601211401 & 46.81 & 55139 & 13.0611 & -73.3214 &
		\begin{tabular}[c]{@{}c@{}}15.52$\pm$1.83\end{tabular} &\begin{tabular}[c]{@{}c@{}}7.14$\pm$0.84\end{tabular}& ... & \begin{tabular}[c]{@{}c@{}}\end{tabular}
		
		\\ \hline
		\textbf{Suzaku}   & \textbf{} & & & &\textbf(0.5-7 keV)&&&\\ \hline
		 503094010 & 120.48 & 54629 & 13.0583 & -73.3219 & 48.41$\pm$45.65 & 22.26$\pm$20.99 & ... \\
		\hline
		\textbf{NuSTAR} & \textbf{FPM-A} & & & &\textbf(3-10 keV)&&& \\
		\hline
		50311002002 & 169.75  & 57824 & 13.0612 & -73.3209 & 211.56$\pm$15.3 & 97.32$\pm$7.04  &\textbf{305.49} \\
		50311002004 & 76.77 & 57953 & 13.0548 & -73.3197 &  2110$\pm$352 &970.66$\pm$161.93 &15.28 \\
		50311002006 &29.94 &57974 & 13.0552 & -73.3218 &...  &... & ...  \\
		30361002003 & 0.31&58418 & 13.0545 & -73.3219 &...  &... & ...   \\
		30361002004 &   58.35& 58421 & 13.0545 & -73.3209 & 2899$\pm$233 & 1333.62$\pm$107.18 &15.24  \\
		30361002002 & 70.65  & 58087 & 13.0535 & -73.3230 &  5136$\pm$94 &2362.7$\pm$43.24 &15.26  \\
		\hline
	\end{tabular}\par
	\end{centering}
	\medskip
	Table displaying Pulsations and other X-ray data for all sources falling in the FoV of either of the two pulsars SXP15.3 and SXP305, from Chandra ACIS, XMM-Newton EPIC, Suzaku and NuSTAR(FPM-A). SXP305 was found pulsating by \textit{NuSTAR} while SXP15.3 pulsations were detected once by \textit{Chandra} and thrice by \textit{NuSTAR}. The \textit{Chandra} rows which are empty corresponds to the ACIS-S observations where the targeted source lay outside the chip edges. For each of the individual \textit{Chandra} observations, the first sub row in the flux column refers to the flux obtained about the SXP15.3 coordinate\citep{Coe&Kirk2015} and the next entry corresponds to the flux obtained about the SXP305\citep{2019ApJ...884....2L} coordinate. 
	\end{small}    
\end{table*}

\begin{table*}
\begin{small}
	\begin{centering}
	\caption{}
	\label{tab:RXTE_obs}
	\begin{tabular}{cccccccc}
		\hline
		\textbf{Obs. ID} & \textbf{Time} &\textbf{MJD}  & \textbf{RA} & \textbf{Dec}  & \textbf{Flux} & \textbf{L\textsubscript{x}}/$10^{34}$ &  \textbf{Period} \\
		 & (ks) & (d) & (deg)  & (deg) & (count~PCU$^{-1}~s^{-1}$)  &(erg~s$^{-1}$) & (s) \\
		\hline
		\textbf{RXTE} & \textbf{PCA} & & & &\textbf(keV)& & \\
		\hline
		
		50088-02-01 & 7.37 & 51918 &  12.4999 & -73.0999  &  &  & ...  \\
		     &      &       &         &        &          0.32     &  1.8            & 306.74$\pm$5.41         \\
		\hdashline
		92076-04-42 & 8.98 & 54242 &  12.4999 & -73.0999 &  &  & ...  \\
		 &      &       &           &         &   0.17  &     1.0          & 307.69$\pm$3.19       \\
		\hdashline
		93037-02-14 & 6.90 & 54380 & 13.3999 & -73.3999 &  &  & ... \\
		 &      &       &           &         &       0.5        & 2.8               & 301.20$\pm$8.62         \\
		 \hdashline
		 	93037-02-15 & 7.85 & 54391 & 13.3999 & -73.3999 &  &  & ...  \\
		     &      &       &          &         &    0.25           &        1.4       & 303.95$\pm$4.15        \\
		\hdashline
		93037-02-29 & 7.42 & 54491 & 13.3999 & -73.3999 &  &  & ...  \\
		 &      &       &           &         &  0.24   &         1.35      & 301.20$\pm$4.08         \\
		\hdashline
		94037-04-41 & 9.30 & 55153 & 12.4999 & -73.0999 &  &  & ... \\
		 &      &       &           &         &   0.21            &   1.2            & 304.87$\pm$4.64         \\
		 \hdashline
		 95037-04-22 & 6.88 & 55345 & 12.4999 & -73.0999 & &  & ...  \\
		     &      &       &           &         &   0.28            &   1.6           & 303.03$\pm$2.06        \\
		\hdashline
		95037-04-48 & 12.37 & 55532 & 12.4999 &  -73.0999 &  &  & ...  \\
		 &      &       &           &         &   0.17  &     1.0          & 307.69$\pm$3.31         \\
		\hdashline
		95037-04-52 & 10.70 & 55556 & 12.4999 & -73.0999 & 0.11 &  & 15.24 \\
		 &      &       &           &         &   0.19            &  1             & 306.74$\pm$2.82        \\
		 \hdashline
		 96037-04-25 & 11.13 & 55732 & 12.4999 & -73.0999 &  &  & ...\\
		 &      &       &           &         &   0.25            &   1.4            & 308.64$\pm$3.33 \\       \\
		\hline
		\hline
	\end{tabular}\par
	\end{centering}
	\medskip
	Table displaying all \textbf{RXTE-PCA} candidate detections of SXP305 around P=305 secs (n=1). The exposure times are in ksecs and the RA and Dec are the pointing RA and Dec of the telescope for that ObsID. There was a single ObsID: 95037-04-22 for which both the pulsars SXP15.3 as well as SXP305 showed pulsations.
	\end{small}    
\end{table*}


\bsp	
\label{lastpage}
\end{document}